\documentclass{aa}
\usepackage{graphics}
\begin{document}
\thesaurus{03 (11.01.1; 11.19.5; 11.19.6; 11.14.1; 11.09.1 M 31)}
\title{The stellar population of the decoupled nucleus in M~31}
\author{O. K. Sil'chenko
\inst{1,2}
\and A. N. Burenkov
\inst{3,4}
\and V. V. Vlasyuk
\inst{3}
}
\offprints{O. K. Sil'chenko}
\institute{
Sternberg Astronomical Institute, University av. 13,
Moscow 119899, Russia
\and
Isaac Newton Institute, Chile, Moscow Branch
\and
Special Astrophysical Observatory, Nizhnij Arkhyz, 357147 Russia
\and
RGO Astronomy Data Centre, Guest Investigator}
\date{Received <date>/ Accepted <date>}
\maketitle

\begin{abstract}

The results of a spectroscopic and photometric investigation of the
central region of M~31 are presented. An analysis of absorption-index
radial profiles involving magnesium, calcium, and iron lines
has shown that the unresolved nucleus of M~31 is distinct by its
increased metallicity; unexpectedly, among two nuclei of M~31,
it is the faintest one located exactly in the dynamical center
of the galaxy (and dynamically decoupled) which is
chemically distinct. The Balmer absorption line $H\beta$ has been
included into the analysis to disentangle metallicity and age
effects; an age difference by a factor 3 is detected between
stellar populations of the nucleus and of the bulge, the nucleus
being younger. The morphological analysis of CCD images has revealed
the presence of a nuclear stellar-gaseous disk with a radius of some
100 pc, the gas component of which looks non-stationary, well
inside the bulge of M~31.

\keywords{Galaxies: M~31; nuclei; structure; stellar content;
          abundances}

\end{abstract}

\section{Introduction}

The stellar core of the nearby spiral galaxy \object{M~31} is
decoupled both dynamically and chemically from the bulge of
the galaxy.

Discussion on the dynamical distinctness of the nucleus of
\object{M~31} began in 1988 when Dressler and Richstone
(\cite{dr88}) and Kormendy (\cite{k88}) published kinematical data
for stars in the center of
\object{M~31}: the rotation curve reached a sharp maximum at a radius
of 1\arcsec\ followed by a drop practically toward zero velocity;
the stellar velocity dispersion had also a prominent maximum in the
center. The dynamical distinctness of the nucleus of \object{M~31}
was then explained by the presence of a supermassive black
hole. Later the black hole hypothesis was confirmed when an
image obtained with the HST revealed
that a mass of $3 \,\ 10^7\, M_{\sun}$ attached to the dynamically
decoupled nucleus is concentrated not in the bright star-like source
P1 but in a fainter nucleus P2, $0\farcs5$ from P1
(\cite{letal93}). The true nucleus of \object{M~31}, P2,
is quite faint optical source, the mass-to-luminosity ratio is high;
so the idea of the presence of a supermassive black hole in the center
of \object{M~31} is finally confirmed. However even proponents of the
supermassive black hole hypothesis accepted the simultaneous presence
of a dynamically decoupled stellar subsystem, namely, of a compact
nuclear disk with a radius of 3\arcsec--5\arcsec\ (10--17 pc).
Kormendy (\cite{k88}) found a zone of low stellar velocity
dispersion in the radius range $1\arcsec < r < 4\arcsec$, that is,
a cold nuclear stellar subsystem embedded into the bulge, and
Tremaine (\cite{tr}) had claimed from dynamical arguments that
the "double" nucleus in \object{M~31} can be stable
during several billion years as a
thick eccentric elliptical nuclear disk which contained the
supermassive black hole at one of its foci. So the presence
of a \emph{stellar} dynamically decoupled nucleus -- or compact
disk -- in the center of \object{M~31} is also widely accepted.

Interestingly, the chemical distinctness of the nucleus
of \object{M~31}
was noted some decades earlier than its dynamical distinctness,
at the romantic epoch when one photographic spectrum of the
galaxy had to be exposed during three nights. Joly and Andrillat
(\cite{ja73}) pointed out a change of the \ion{Na}{i}, \ion{Mg}{i},
CN, and \ion{Ca}{ii}K
equivalent widths between the nucleus and the bulge; if compared
to galactic globular clusters, the magnesium and calcium abundance
of the bulge is --0.84 dex, and the nucleus
is more metal-rich by 0.6 dex in calcium and by 1.5 dex
in magnesium. Surprisingly, the data of Joly and Andrillat
(\cite{ja73}) on the iron absorption lines show a prominent
equivalent width gradient in the bulge, but the nucleus seems
to share this gradient and does not look distinct. A CN break
between the
nucleus and the bulge of \object{M~31} was also noted by McClure
(\cite{mc}) though not by spectroscopy but by narrow-band
photometry: filters with passbands of 83--85 \AA\ were centered
on the absorption line CN$\lambda 4165$ and on the off-line continuum
at $\lambda 4255$. Spinrad and Liebert (\cite{sl75}) found another
photometric change, $\Delta (U-V) = 0.13 \pm 0.04$; but this colour
change corresponds to a very modest metallicity break, not more than
0.2 dex. Morton and Andereck (\cite{ma76}) have confirmed the result
of Joly and Andrillat (\cite{ja73}) on the equivalent width change
for \ion{Ca}{ii}K line -- $EW_{nuc}/EW_{bulge}=1.38 \pm 0.11$, --
and Bica et al. (\cite{basch90}) have also noted changes in the
\ion{Mg}{i} and \ion{Ca}{ii}
lines, but not so prominent as those in the paper of Joly and
Andrillat (\cite{ja73}): if calibrated into metallicity,
they gives a metallicity difference of 0.2 dex between the
nucleus and the bulge. Cohen (\cite{cohen}) exposed a long-slit
spectrum of \object{M~31} along the east-west direction by using
a linear digital detector; she found changes of the
absorption lines NaD and Mgb and, by modelling them with stellar
population synthesis, obtained a metallicity difference of
four times (0.6 dex). In general, though the existence of metallicity
difference between the nucleus and the bulge of \object{M~31} has been
established long ago, quantitative estimates
range from 0.2 dex to an order of magnitude. Moreover, the studies of
abundances of the stellar populations in the center of \object{M~31}
in the 60--70 was followed by a long pause, and
investigations with modern CCDs have not been
undertaken. The only exception is the work of Davidge (\cite{dav97});
but he does not concern the nucleus being concentrated on gradients
of absorption-line equivalent widths in the inner bulge. Another,
even more recent work of Davidge et al. (\cite{detal97}) treats the
properties of the stellar populations in the nucleus of \object{M~31}
based completely on photometric data; their C-M diagrams for
individual stars being constructed under very good seeing conditions
($FWHM \approx 0\farcs15$) in the narrow concentric rings with
radii of $0 - 1\farcs4$, $1\farcs4 - 2\farcs8$,
and $2\farcs8 - 4\farcs0$ have allowed to detect a noticeable
increase in mean stellar age with radius in the innermost part
of the galaxy.

Our paper presents a spectral investigation of the stellar population
properties in the nucleus and in the inner bulge of \object{M~31}.
In contrast to our precursors, we have now the possibility to compare
both the mean metallicity and the mean age of stars and to determine
in this way a sequence of star formation epochs. The work is
undertaken within a large observational project on
chemically distinct nuclei in galaxies and on
possible relations between chemically and dynamically
decoupled galactic nuclei.

\section{Observations and data reduction}

The observations of \object{M~31} were performed in September 1996,
at the 1\,m telescope of the Special Astrophysical Observatory of
the Russian Academy of Sciences (Nizhnij Arkhyz, Russia) with the
long-slit spectrograph UAGS equipped by a $1040 \times 1160$ pixel
CCD. The slit width was $3\farcs2$, the seeing quality was
similar, about of 4\arcsec. In the night of September 11/12 the
galaxy was observed in three position angles, 35\degr, 80\degr, and
125\degr, with total exposure times of 40 minutes for each position
angle, and in the night of September 12/13 -- in two more position
angles, 155\degr\ and --10\degr, with a total exposure time of 30
minutes for each. Each time a blank sky area in 2\degr\ from the
center of \object{M~31} was exposed; after that the sky spectra
were smoothed and subtracted from the spectra of the galaxy. During
the observations we used a grating of 651 grooves per mm which
provided a dispersion of 1.59\,\AA/px and a spectral resolution of
3.5--4\,\AA; the spectral range was 3900--5700 \AA.
The scale along the slit was $1\farcs54$/px, and as the seeing
quality was bad, we binned by 3 rows and studied absorption-line
equivalent width variations along the slit with a step of $4\farcs6$.

Besides our own observations we have used spectral data for
\object{M~31} obtained from the La Palma Archive. The galaxy was
observed at the 4.2\,m William Herschel Telescope on September 19,
1991, with the two-armed long-slit spectrograph ISIS equipped with
a $800 \times 1180$ pixel CCD. The total exposure
time was 1.5 hour. The galaxy was exposed in one position angle,
$P.A.=55\degr$, with a slit width of $0\farcs7$. The dispersion was
0.74\,\AA/px (spectral resolution of 2\,\AA), the blue-arm spectral
range which is more interesting for us expanded over
3800--4700\,\AA. The scale along the slit was $0\farcs34$/px.
The seeing was $0\farcs9$.

\begin{figure}
\resizebox{\hsize}{!}{\includegraphics{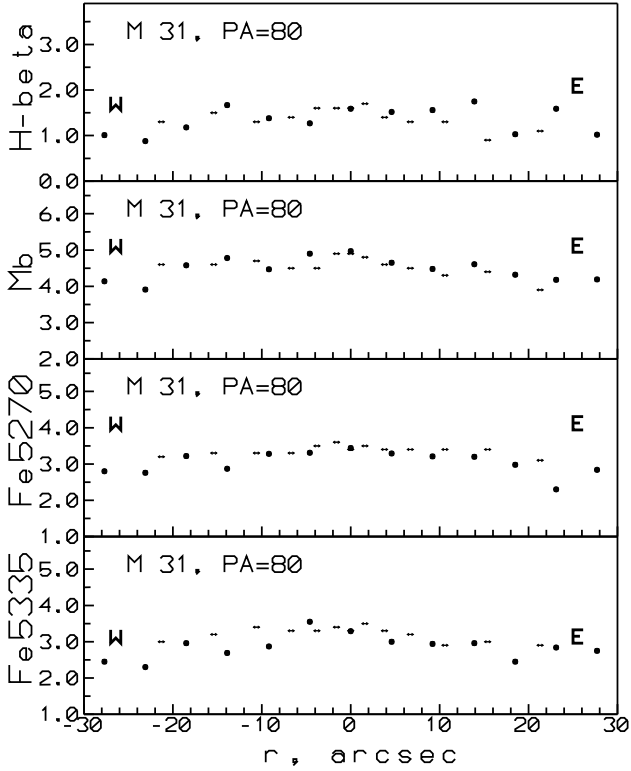}}
\caption{Radial profiles of absorption-line indices $H\beta$,
   Mgb, Fe5270, and Fe5335 in $P.A.=80\degr$ (points) compared to
   the data of Davidge (1997) in $P.A.=90\degr$
   (two-sided arrows)}
\label{comp}
\end{figure}

To reduce the spectral data -- remove cosmic-ray hits, subtract bias,
extract one-dimensional spectra by taking into account the geometry of
the images, calibrate wavelengths and linearize the spectra -- we
have used a software developped by one of us (\cite{vlas}). After that
we calculated absorption-line indices for the strong lines of calcium,
iron, magnesium, and hydrogen (for definition of the indices see
Worthey et al. \cite{wfgb94} and Worthey \& Ottaviani \cite{wo97})
and constructed profiles of the indices along
the slit. We have used the spectrum of \object{M~31} obtained at the
WHT to calculate the indices Ca4227, $H\delta A$,
$H\gamma A$, Fe4383, Ca4455, and Fe4531, and our spectra to
calculate the indices $H\beta$, Mgb, Fe5270, and Fe5335 missed
in the spectral
range of the ISIS blue arm. We did not observe standard Lick stars,
but our instrumental index system for $H\beta$, Mgb, Fe5270,
and Fe5335 is close to the Lick system as the spectral resolution is
similar. To demonstrate this fact, we compare our index
measurements in $P.A.=80\degr$ with the data of Davidge (\cite{dav97})
whose slit was set in the direction of east-west (Fig.~\ref{comp}).
Davidge (\cite{dav97}) observed standard stars and transformed his
measurements into the Lick system; he needed this procedure because
of his spectral resolution of 14 \AA. We see in Fig.~\ref{comp} that
an agreement is excellent inside the claimed accuracy of Davidge's
measurements, 0.2 \AA\ for $H\beta$, 0.3 \AA\ for Mgb, 0.1 \AA\
for Fe5270, and 0.2 \AA\ for Fe5335.

By combining the indices of different
chemical elements and by comparing them with models, we are able
to reach some conclusions about the stellar population properties
at various distances from the center of \object{M~31}.

\section{Results}

\subsection{Chemical and age decoupling of the nucleus of M~31}

\begin{figure}
\resizebox{\hsize}{!}{\includegraphics{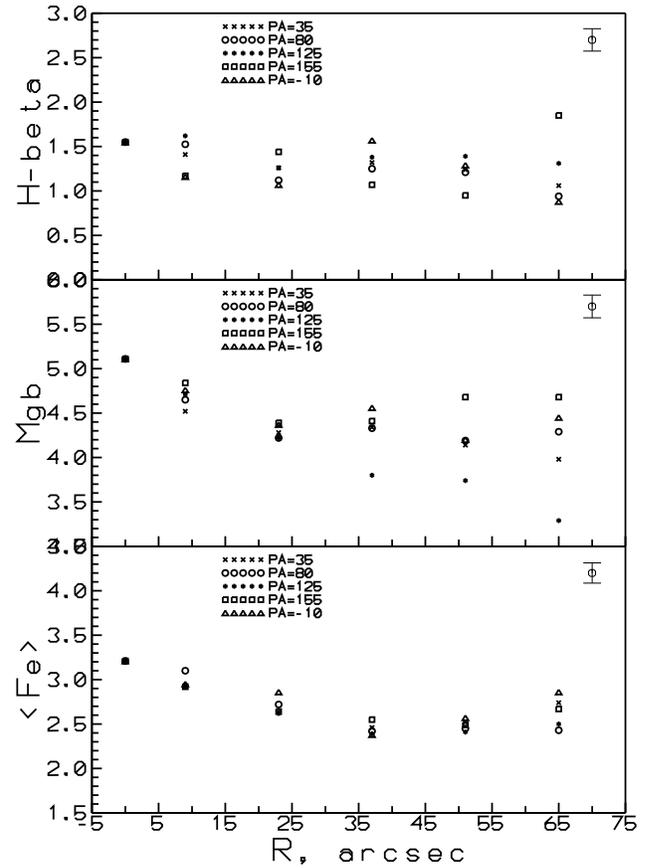}}
\caption{Radial profiles of absorption-line indices $H\beta$,
   Mgb, and $<\mbox{Fe}>$ in five position angles, averaged over
   radius intervals
   of 14\arcsec. Typical error bars are shown in the upper right
   corners of the frames; they are somewhat smaller for measurements
   closer to the nucleus and somewhat larger at the edge}
\label{prof1m}
\end{figure}

By using the five cross-sections of \object{M~31} obtained by us
at the 1\,m telescope of SAO RAS, we have calculated index profiles
for $H\beta$, Mgb, Fe5270, and Fe5335 up to 70\arcsec\ from
the center with a step of 14\arcsec. Two iron indices, Fe5270 and
Fe5335, were merged into
$<\mbox{Fe}> \equiv (\mbox{Fe5270+Fe5355})/2$.
The two halves of each profile, symmetric around the nucleus, were
averaged. The results are presented in Fig.~\ref{prof1m}, the error
bars being estimated by point-to-point scatter under binning of
$4\farcs6$.

In accordance with numerous results of earlier investigations, the
center of \object{M~31} is remarkable by the higher equivalent widths
of magnesium and iron lines. The measurements at $r=9\arcsec$ may be
affected by seeing (let us remind that the seeing during our
observations was not better than 4\arcsec). At larger distances from
the center, in the radius range of 23\arcsec--65\arcsec, radial
index gradients if they exist are negligible with respect to the index
changes between the nucleus and the bulge. The mean index values in
the bulge averaged over all the five cross-sections in the
radius range 23\arcsec--65\arcsec\ are
Mgb(bulge)=$4.25\pm 0.02$ and
$<\mbox{Fe}>(\mbox{bulge})=2.56 \pm 0.01$.
If we approximate the radial index dependencies by linear laws and
extrapolate them to $r=0$, we would obtain central bulge index values
Mgb(bulge)=$4.37\pm 0.06$ and
$<\mbox{Fe}>(\mbox{bulge})=2.68\pm 0.03$. For the
nucleus we have measured
Mgb(nuc)=$5.11\pm 0.06$ and $<\mbox{Fe}>(\mbox{nuc})=3.21\pm 0.05$.
Therefore, the differences between the nucleus and the bulge are
$\Delta \mbox{Mgb}=0.86\pm 0.10$ and $\Delta <\mbox{Fe}>=0.65\pm 0.06$
(or $\Delta \mbox{Mgb}=0.74\pm 0.12$ and
$\Delta <\mbox{Fe}>=0.53\pm 0.08$,
if we use linear index radial dependencies for the bulge).
Let us note that due to strongly increased stellar velocity
dispersion in the nucleus of \object{M~31} the absorption
lines there may be broadened out of the index measuring ranges;
so the nuclear indices may be underestimated, and the real
index differences between the nucleus and the bulge may be
even larger. Application of models of Worthey (\cite{w94})
for an old stellar population under the
assumption of equal bulge and nucleus ages and of a solar
[Mg/Fe]=0 gives an estimate of metallicity difference,
$0.42 \pm 0.05$ dex (or $0.35\pm 0.06$ dex with a linear gradient
in the bulge), the same for
magnesium and iron. But are the assumptions of
equal ages for the nucleus and bulge stellar populations and of
solar magnesium-to-iron ratio valid in this particular case?

\begin{figure}
\resizebox{\hsize}{!}{\includegraphics{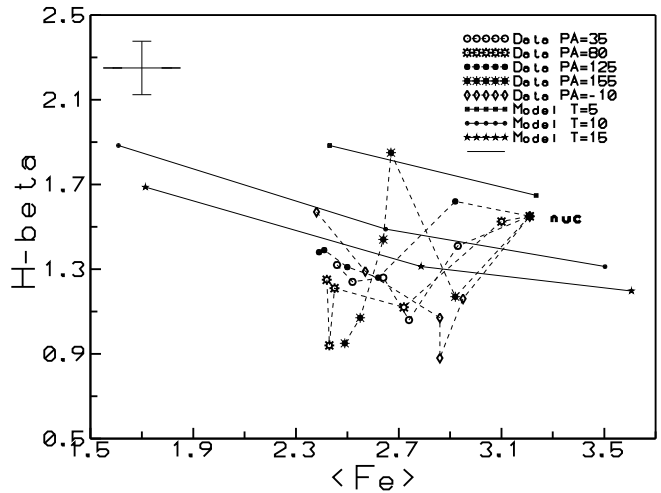}}
\caption{The diagram ($H\beta$, $<\mbox{Fe}>$) for our data
       on M~31
       in five position angles; the nucleus is marked by 'nuc',
       other points along the radius are connected by dashed
       lines. The solid lines present the models of Tantalo
       et al. (1998) for [Mg/Fe]=+0.3, the ages of the
       models are given in billion years. The typical error bar
       for our measurements is shown in the upper left corner
       of the frame}
\label{diag}
\end{figure}

If the nucleus is chemically decoupled, it would be natural to suggest
that the epochs of basic star formation in the nucleus and in the
bulge are different. So the mean ages of the stellar populations in
the nucleus and in the bulge are expected to be different. It is known
that the hydrogen absorption lines are much more sensitive to age
than to metallicity, so by confronting a Balmer line index
with a magnesium or iron index one can disentangle age and metallicity
effects. Such attempts were undertaken for example by
Worthey (\cite{w94}) and Worthey and Ottaviani (\cite{wo97}) in the
case of a fixed solar magnesium-to-iron ratio. Recently Tantalo
et al. (\cite{tcb98}) have presented new model calculations and have
written a system of three linear equations allowing to determine
differences in metallicity, age, and magnesium-to-iron ratio from
the $H\beta$, $\mbox{Mg}_2$, and $<\mbox{Fe}>$ differences.
We have taken the
index differences between the nucleus of \object{M~31} and the bulge
at $r=23\arcsec$ (Fig.~\ref{prof1m}), have used the relation
Mgb $\approx 15 \mbox{Mg}_2$ (Worthey \cite{w94}), and from the
equations of Tantalo et al. (\cite{tcb98}) we have derived
differences of the parameters:
$\Delta \mbox{[Mg/Fe]}=+0.12$, $\Delta \log (Z/Z_\odot)=+0.53$
and $\Delta \log t=-0.52$. This means that the magnesium-to-iron
ratios are close and the nucleus is three times more metal-rich and
three times
younger than the bulge at 23\arcsec\ from the center. As for absolute
values, the model of Tantalo et al. (\cite{tcb98}) which gives the
best set for the index combination in the nucleus corresponds to
$Z=0.05$ (or $2.5Z_{\sun}$), [Mg/Fe]=+0.3 and $t=6-7$ billion years.
Consequently, in the bulge the stellar population metallicity is
slightly below the solar one, and its age is over 15 billion years.
Figure~\ref{diag} presents the combination of a theoretical diagram
($H\beta$, $<\mbox{Fe}>$) for [Mg/Fe]=+0.3 from the work
of Tantalo et al. (\cite{tcb98}) and of our data for \object{M~31}
from Fig.~\ref{prof1m}. One can see immediately that the locations of
the nucleus and of the bulge measurements in the diagram
($H\beta$, $<\mbox{Fe}>$) imply a significant difference of
stellar population mean ages: the nucleus has an age about of 7
billion years, and the bulge measurements over the whole radius range
of 23\arcsec--65\arcsec\ are below the model sequence of 15 billion
years, so the bulge stellar populations are everywhere older than
15 billion years. One danger always exists when one makes such
an analysis: the $H\beta$ absorption line equivalent width
may be affected by emission. We know (see e. g.
Ciardullo et al., \cite{ciar}) that at $R > 5\arcsec$ emission
lines are seen in the spectra of \object{M~31}. Unfortunately,
we have not found quantitative estimates of their equivalent
widths in the literature. But a visual analysis of the Figs.~7 and 8
in the paper of Ciardullo et al. (\cite{ciar}) allows us to
estimate roughly that the equivalent width of the $H\alpha$
emission line is less than 1 \AA\ in the radius range of
20\arcsec -- 60\arcsec. It means that a correction for the
emission which must be applied to the $H\beta$ absorption
indices in Fig.~\ref{diag} is less than 0.3 \AA, and therefore
the estimate of the bulge stellar population mean age still
remains larger than 10 billion years, and the age difference
between the nucleus and the bulge remains substantial.

\begin{figure}
\resizebox{\hsize}{!}{\includegraphics{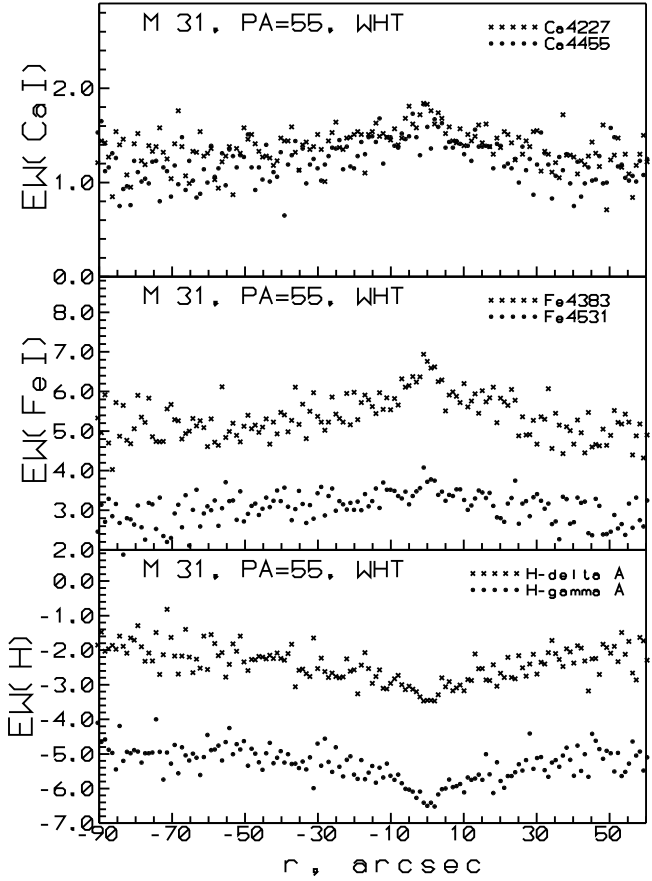}}
\caption{Radial profiles of calcium, iron, and hydrogen
   absorption-line indices according to the WHT data,
   averaged over radius intervals of 1\arcsec}
\label{profwht}
\end{figure}

\object{M~31} is known to have two nuclei (\cite{letal93}).
Therefore it would be important to localize the chemically
distinct entity more exactly. For this purpose we have used
the long-slit data from the 4.2\,m WHT obtained under much better
seeing conditions than our observations. Figure~\ref{profwht}
presents the absorption-line index variations along the slit for
\ion{Ca}{i} (Ca4227 and Ca4455), \ion{Fe}{i} (Fe4383 and Fe4531)
and hydrogen ($H\gamma A$ and $H\delta A$) in the
position angle $P.A.=55\degr$, very close to the line connecting two
nuclei of \object{M~31} ($43\degr \pm 1\degr$, \cite{letal93}).

\begin{figure}
\resizebox{\hsize}{!}{\includegraphics{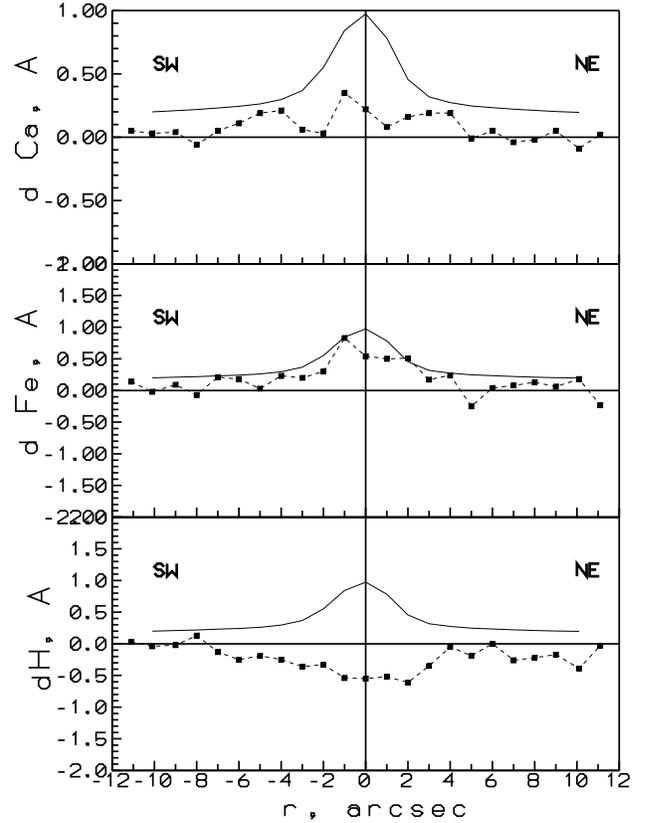}}
\caption{Radial profiles of
   $(\Delta \mbox{Ca4227}+\Delta \mbox{Ca4455})/2$,
   $(\Delta \mbox{Fe4383} + \Delta \mbox{Fe4531})/2$, and
   $(\Delta H\gamma A + \Delta H\delta A)/2$
   indices after removing linear radial trends;
   the solid line shows continuum profile near
   $\lambda \approx 4400\,\AA$, the positions of the brighest
   continuum points are marked by vertical lines}
\label{difprof}
\end{figure}

Again we can see an outstanding nucleus with increased metal-line
indices and decreased hydrogen indices and also a slight index drift
along the radius. The dependencies of indices on $r$ look like linear
ones, so we have approximated them by linear formulae in the radius
range of 5\arcsec--60\arcsec, have extrapolated the formulae to $r=0$,
and have subtracted the fitted linear laws from the measured profiles.
The results of this subtraction averaged over the indices of every
element are presented in Fig.~\ref{difprof}
together with the continuum profile along the slit (the continuum is
taken around $\lambda \approx 4400\,\AA$); only the central part of
the profiles, $r \le 12\arcsec$, is shown.

Both the metal-index profiles exhibit maxima shifted to
the south-west with respect to the continuum maximum. The continuum
peak marks the position of the brighter nucleus, P1, which is not the
isophote center nor the dynamical center. The continuum profile looks
asymmetric, and a Gauss analysis reveals a presence of two components,
a point-like one at $r=0$ and an extended one ($FWHM=7\arcsec$)
at $r=-0\farcs4$. It is the latter, the southern-western component
that must contain the second, fainter
nucleus of \object{M~31}, P2, which corresponds to the photometric
and the dynamical centers of the galaxy (\cite{letal93}).
Therefore we conclude that P2 is
the chemically and age decoupled nucleus of \object{M~31}
(though due to the flatness of the Balmer line index minima inside
$r < 5\arcsec$ we may suspect that the nuclear disk mentioned
by Kormendy (\cite{k88}) has also a metal abundance enhancement).
Let us remind that P2 is dynamically decoupled
(Kormendy \cite{k88}). This result, though
impressive, was not unexpected. King et al. (\cite{ksc95})
obtained an image of \object{M~31} at $\lambda = 1750\,\AA$; they
had found that P2 is brighter in the ultraviolet than P1,
in contrast to the visible. They have immediately
interpreted this fact as an evidence for P2 metallicity
overabundance because there exists a correlation between UV excess
and metallicity in elliptical galaxies. However, we must note that
Davidge et al. (\cite{detal97}) have found that P1 and P2
have roughly equal colours $H-K$; but the formal error
estimates, 0.1--0.2 mag, can hide a metallicity difference as large
as a factor 10 (Worthey \cite{w94}).

\subsection{Morphological analysis of the central part of M~31}

\begin{figure}
\resizebox{\hsize}{!}{\includegraphics{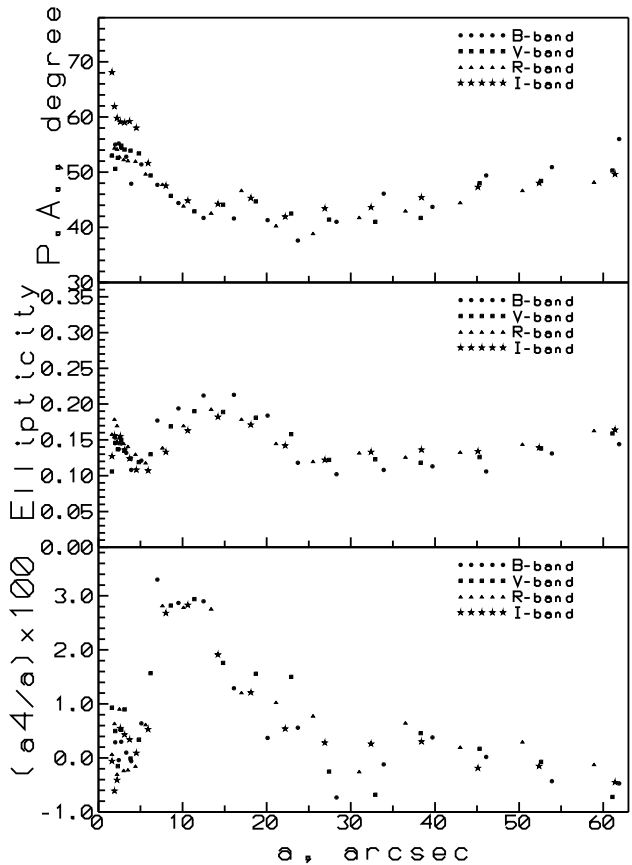}}
\caption{Radial variations of isophote parameters}
\label{isopar}
\end{figure}

We have used CCD images of the central part of \object{M~31} obtained
from the La Palma Archive to analyze the two-dimensional
surface brightness distribution. The galaxy was observed on November
17, 1990,  at the 1\,m Jacobus Kapteyn Telescope through four filters,
$BVRI$, by a $320 \times 512$ pixel RCA CCD, with a seeing quality of
$1\farcs5$. After bias subtraction and division by flat fields,
the isophotes of the images were approximated by ellipses.
Figure~\ref{isopar} presents the radial variations of
the major axis position angles, ellipticities,
and fourth Fourier coefficients $a_4/a$.

\begin{figure}
\resizebox{\hsize}{!}{\includegraphics{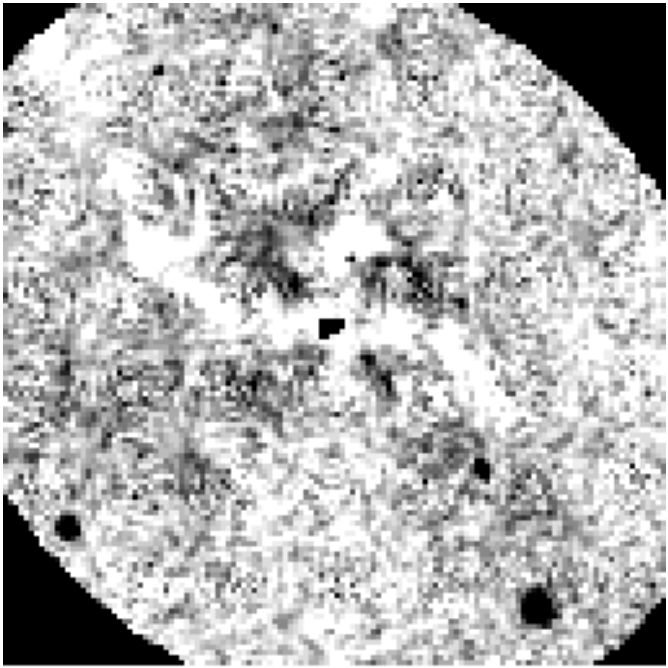}}
\caption{Map of photometric residuals (after pure elliptical
   isophote subtraction) in the filter $R$. The white spots are
   in reality dark ones. North is up, East is to the left, the
   picture sizes are $105\arcsec \times 105\arcsec$}
\label{resid}
\end{figure}

Figure~\ref{resid} is a "residual map",
demonstrating the difference between the observed $R$ image and
a modelled one consisting of pure elliptical isophotes.
The "residual maps" in all four filters are similar and look like
the Fig.~3b from the paper of Wirth et al. (\cite{wsb85}) though
Wirth et al. subtracted median-smoothed images, not modelled ones.
Wirth et al. (\cite{wsb85}) have interpreted the spiral absorption
features of their  Fig.~3b as dust arms. We reach the same conclusion.

Figure~\ref{isopar} is fully consistent with earlier photometric
results for the central part of \object{M~31} (see, for example,
CFHT data in the work of Bacon et al. \cite{bemn94}), though we would
like to note that boxiness of the inner isophotes of \object{M~31}
is shown here for the first time. There are also some
details of the radial profiles of $P.A.$ and $(1-b/a)$ which were
not discussed earlier. Over all the four spectral bands the major
semi-axis range of 7\arcsec--25\arcsec\ is characterized by a local
maximum of ellipticity which reaches 0.2 and by a local minimum
in the position angle which oscillates between 40\degr\ and 44\degr\
close enough to the orientation of the line of nodes of the disk,
$P.A.=38\degr$. The same radius interval is characterized by a
$a_4/a$ higher than +1\%; outside, $a_4/a$ is less than +1\%.
This means that in the radius range of 7\arcsec--25\arcsec\ we see
a stellar disk whose plane is close to or even coincides with the
global plane of the galaxy. One must keep in mind that the central
region of \object{M~31} is photometrically dominated
by the bulge (\cite{jap76}). At the distance of \object{M~31}
the major semi-axis of 25\arcsec\ corresponds to a metric radius
of 85 pc, therefore we deal with a so called nuclear disk; similar
disks were earlier found in some other early-type spiral galaxies,
particularly, in \object{NGC~4594} (\cite{burk}). In
Fig.~\ref{resid} one can see two dark (dust?) spiral miniarms
located also in the same radius range. Besides that, over the same
radius range a noticeable emission line [OIII]$\lambda 5007$ is
seen in our spectra (in the very center of \object{M~31},
$r < 5\arcsec$, emission lines are absent according to the claims of
Bacon et al. \cite{bemn94}). We can conclude that the nuclear
stellar disk of \object{M~31} contains also dust and ionized gas.
Ionization might be due to shocks because, according to
Ciardullo et al. (\cite{ciar}), the nitrogen emission line
[NII]$\lambda 6583$ is everywhere stronger than $H\alpha$
inside $r=1\arcmin$.

\section{Discussion and conclusions}

Though \object{M~31}, the most nearby spiral galaxy, has been studied
for a long time and with unmatched spatial resolution, the true
structure of its central region is still a puzzle. The large amount
of accumulated observational data has led to reject many models, even
regarded earlier as reasonable ones. Our work
contributes to this process.

Two extraordinary features -- the strong increase of angular rotation
velocity and of stellar velocity dispersion inside a radius of
1\arcsec\ and the shift of the bright point-like nucleus with respect
to the center of the galaxy -- were initially explained in the frames
of two equally successful models: a strong anisotropy (end-on
tumbling minibar: \cite{gerhard}) and an axisymmetric potential
distribution with a supermassive black hole in the center (Kormendy
\cite{k88}). The former model looked preferable because of the
isophote twist along the radius of \object{M~31} and because of a
noticeable line-of-sight velocity gradient along the photometric
minor axis (Ciardullo et al. \cite{ciar}) -- these observational facts
had allowed Stark and Binney (\cite{sb94}) to suggest the presence of
a large-scale bar in \object{M~31} extended up to 3\arcmin\ from the
center. However Bacon et al. (\cite{bemn94}) observed a central part,
$r \le 5\arcsec$, of \object{M~31} with the Integral Field Spectrograph
TIGER and could compare the surface brightness map with the
line-of-sight velocity field; they have found a coincidence of the
dynamical and photometric major axes and in this way proved the
circularity of stellar rotation inside a radius of 5\arcsec. The
nuclear region of \object{M~31} has appeared to be axisymmetric.
After that a model of Tremaine (\cite{tr}) has become very popular:
he has proposed an eccentric disk rotating in accordance with the
Kepler's law and having a supermassive black hole in one of its
foci, P2. In the frame of this model the bright nucleus P1 is
an apocenter where stars decelerate and their orbits crowd, then
forming a surface brightness excess. The model of Tremaine (\cite{tr})
is often cited as the most realistic; however it has been dismissed
two years ago. Gerssen et al. (\cite{gkm95})
had observed \object{M~31} with a long-slit spectrograph; they had
oriented the slit along the minor axis of the nucleus
($P.A.=148\degr$) and set its width as $1\farcs25$, so that the
putative disk of Tremaine falls completely within the slit. Due to
angular momentum conservation, the luminosity-weighted velocity
of a filled Keplerian orbit about a stationary object should be
zero; but the analysis of the observations showed the presence of
two distinct kinematical components at $r=0$! Indeed, only one model
is not rejected: that of a giant
stellar cluster falling to the center of \object{M~31} under the
attraction of a black hole (\cite{emcomb97});
but such a configuration is unstable and very short-lived, so it
implies that \object{M~31} is observed during an unique
evolutionary stage.

The results obtained in this work even worsen the situation. Earlier
when we discovered a chemically distinct nucleus in galaxies more
distant than \object{M~31}, we supposed it to be a compact nuclear
disk; in several cases where the chemically distinct nuclei appeared
to be resolved, our supposition was confirmed by a coincidence of
chemically, kinematically and photometrically distinct area radii
(\object{NGC~1052} -- \cite{me95}, \object{NGC~4621} -- \cite{me97};
but note that they are elliptical galaxies). The case of \object{M~31}
has destroyed our presumption: the nuclear disk of \object{M~31}
detected by its photometric signature has a radius of 80--100 pc,
while the chemically distinct nucleus is unresolved, being less than
3 pc. Moreover, it is not P1 which is chemically distinct, though
only P1 is thought to be a giant stellar cluster, but P2 which is
assumed to be a supermassive black hole the stellar content around
which is not yet known. The central region with a radius of $2\farcs3$
in Fig.~\ref{diag} looks three times younger than the bulge; but P2
contributes only a small fraction into the luminosity of this region,
namely, 30\%\ if to assume the photometric decomposition model C
from the work of Bacon et al. (\cite{bemn94});
so we must conclude that the stellar population of P2 is significantly
younger than 5 billion years. Hence a relatively recent star formation
burst exactly at the dynamical center of \object{M~31} seems
indubitable. It may not be the only and last one: dust spiral arms in
the nuclear disk of \object{M~31}, the local splitting of emission
lines in several spots within a dozen arcseconds from the center
seen in our data and also in the data of Ciardullo et al. (\cite{ciar}),
the bright nucleus P1 which cannot be on a stable orbit -- all these
facts are evidences in favour of a continuous matter drift to the center
of \object{M~31}. Though less probable, there exists still another
explanation of the observational facts reported by us: if in the
proximity of the supermassive black hole located inside P2 the stars
have some unusual structure, say, if their external atmospheres are
removed by tidal effects, then we see their inner layers which are
enriched by metals. In this case the nucleus P2 may look like a
chemically decoupled nucleus too.

\begin{acknowledgements}
This research has made use of the La Palma Archive. The telescopes
WHT and JKT are operated on the island of La Palma by the Royal
Greenwich Observatory in the Spanish Observatorio del Roque de los
Muchachos of the Instituto de Astrofisica de Canarias. The work
is supported by the grant of the Russian Foundation for Basic
Researches 98-02-16196 and by the Russian State Scientific-Technical
Program "Astronomy. Basic Space Researches" (the section "Astronomy").
\end{acknowledgements}

\end{document}